\documentclass{article}

\usepackage{arxiv}

\usepackage[utf8]{inputenc} 
\usepackage[T1]{fontenc}    
\usepackage{hyperref}       
\usepackage{url}            
\usepackage{booktabs}       
\usepackage{amsfonts}       
\usepackage{nicefrac}       
\usepackage{microtype}      
\usepackage{lipsum}
\usepackage{graphicx}
\graphicspath{ {./images/} }

\title{A Serious Game for Simulating Cyberattacks to Teach Cybersecurity}

\author{
 Christopher Scherb \\
  University of Applied Sciences and Arts\\
  Northwestern Switzerland\\
  \texttt{christopher.scherb@fhnw.ch} \\
   \And
 Luc Bryan Heitz \\
University of Applied Sciences and Arts\\
Northwestern Switzerland\\
\texttt{luc.heitz@fhnw.ch} \\
  \And
 Frank Grimberg \\
University of Applied Sciences and Arts\\
Northwestern Switzerland\\
\texttt{frank.grimberg@fhnw.ch} \\
 \And 
  Hermann Grieder \\
 University of Applied Sciences and Arts\\
 Northwestern Switzerland\\
 \texttt{hermann.grieder@fhnw.ch} \\
 \And 
  Marcel Maurer \\
 Muuri Solutions,  \\
 Bern, BE, Switzerland\\
 \texttt{info@muuri.solutions} \\
}

\begin{document}
	
	\maketitle
	
	\begin{abstract}
		With the rising number of cyberattacks, such as ransomware attacks and cyber espionage, educating non-cybersecurity professionals to recognize threats has become more important than ever before. 
		However, traditional training methods, such as phishing awareness campaigns, training videos and assessments have proven to be less effective over time.
		Therefore, it is time to rethink the approach on how to train cyber awareness. In this paper we suggest an alternative approach -- a serious game -- to educate awareness for common cyberattacks. While many serious games for cybersecurity education exist, all follow a very similar approach: showing people the effects of a cyber attack on their own system or company network. For example, one of the main tasks in these games is to sort out phishing mails. 
		We developed and evaluated a new type of cybersecurity game: an attack simulator, which shows the entire setting from a different perspective. Instead of sorting out phishing mails the players should 
		write phishing mails to trick potential victims and use other forms of cyberattacks. Our game explains the intention of each attack and shows the consequences of a successful attack. This way, we hope, players will get a better understanding on how to detect cyberattacks.
		
	\end{abstract}
	
	
	\setcounter{tocdepth}{2}
	
	%
	%
	
	\section{Introduction}
	\label{sect:introduction}
	The past few years have seen a significant rise in the number of cyber attacks across the world. With the increased digitization of business processes, home automation~\cite{scherb2021scoiot}, connected cities~\cite{scherb2018resolution} and the rise of remote work, cyber criminals have found new and sophisticated ways to exploit vulnerabilities in computer networks, computer systems and humans. From phishing scams and ransomware attacks to data breaches and identity theft, cyber attacks have become a major concern for businesses, governments, and individuals alike. The COVID-19 pandemic has also further exacerbated the problem, with cyber criminals taking advantage of the increased online activity to launch targeted attacks on vulnerable individuals and organizations. As technology continues to advance, the threat of cyber attacks is likely to grow, making it imperative for individuals and organizations to stay vigilant and take proactive measures to protect themselves from these digital threats.
	
	Meanwhile, the portrayal of hacking in popular media such as in the series \emph{NCIS}\footnote{\url{https://www.imdb.com/title/tt0364845/}} and the video game \emph{Watch Dogs}\footnote{\url{https://www.ubisoft.com/en-us/game/watch-dogs/watch-dogs}} is often dramatized and unrealistic. In these shows and games, hacking is often depicted as a glamorous and effortless activity, where hackers can break into highly secure systems with just a few keystrokes. While these shows and games may be entertaining, they can perpetuate the misconception that hacking is a harmless activity, leading to an inadequate awareness about the risks associated with cybersecurity. In reality, hacking is a complex and often illegal activity that can result in serious consequences for both the hacker and victim. As the negative impact of hacking, including identity theft, financial fraud, and disruption of critical infrastructure is often inaccurately represented by the media, the media does portray a sense of helplessness once one becomes the target of a hacker. Yet, most attacks could have been avoided by most non-technical employees if they had good security awareness education.
	
	Therefore, one of the most important ways to mitigate the risk of cyberattacks is by educating employees on best practices for cybersecurity~\cite{rahman2020importance, vsvabensky2020cybersecurity}. Employees are often the first line of defense against cyber threats, and without proper training, they may inadvertently expose their company's sensitive information or fall prey to phishing scams~\cite{alkhalil2021phishing}. By providing comprehensive cybersecurity training to employees, businesses can empower them to identify and report suspicious activities, secure their devices and accounts, and adhere to best practices for data protection. This can help prevent costly data breaches and cyber attacks, and also foster a culture of security awareness across the organization. In short, investing in employee cybersecurity education is a crucial step towards safeguarding a company's valuable assets and reputation in today's digital landscape.
	
	However, educating cybersecurity is a quite complex topic and most employees consider training and awareness campaigns as annoying and participate halfhearted. Moreover, recent research has shown, that awareness training such as phishing campaigns have only a short term effect~\cite{lain2022phishing} and do not increase the employees resistance to phishing mails in long term.
	One reason for this is that phishing emails have become increasingly sophisticated, making them more difficult to detect. Attackers use tactics such as social engineering, spoofing legitimate email addresses, and creating convincing fake websites to trick unsuspecting victims. Additionally, some employees may not take phishing threats seriously or may be too busy to fully scrutinize every email they receive. Even with training and awareness programs, employees still may fall for phishing mails. Further, phishing awareness campaigns itself can contribute to naive behavior of employees as dealing with mails from phishing awareness campaigns may not inflict any harm when clicking on arbitrary links. As such, the underestimation of the phishing-related risk may be habituated, making real phishing mails even more critical.
	
	By understanding the tactics used by attackers, such as social engineering and spear-phishing, individuals can be better equipped to recognize and avoid phishing attempts. Additionally, appreciation of the potential consequences of falling victim to phishing can motivate individuals to take the necessary precautions to protect their sensitive information. Many people do not understand what harm clicking on an email or clicking on a link can cause not only to their system but to the entire network, since they have never seen the consequences such as systems encrypted by ransomware, stolen company secrets or financial fraud. 
	
	Therefore, we designed a cybersecurity education game, in which the player experiences cyber attacks from the attacking side~\cite{maurer2022}. The player plays an attacker, who needs to acquire information, sends phishing mails, uses exploits and other forms of cyber attacks to attack a company~\cite{schneier2023hackermind}. The process is presented as realistic as possible. The user has to search for email addresses of victims, fake websites for phishing attacks and buy exploits in a simulated darknet store. The consequences of each attack are shown and explained, so that the player can develop an deeper understanding of the motivation and the different shapes of cyber attacks and by this the resistance against real cyber attacks. 
	
	In the upcoming sections of this paper we will detail our game as follows: Section~\ref{sect:LiteratureReview} contains reviews of literature that introduce the reader to the setting and which contain background knowledge for this paper. Section~\ref{sect:GameDesign} gives an outline of our serious game as a whole and in more detail for each implemented scenario. The results of a short term survey are presented and discussed in Section~\ref{sect:eval}. In Section~\ref{sect:conclusion} we conclude this paper and highlight potential future work.

	
	\section{Literature Review}
	\label{sect:LiteratureReview}
	
	In this section we present and summarize related and background work.
	
	\subsection{ENISA Threat Landscape Report}
	\label{sec:ENISA}
	The ENISA Threat Report\footnote{\url{https://www.enisa.europa.eu/publications/enisa-threat-landscape-2022}}  is an annual publication by the European Union Agency for Cybersecurity (ENISA). The report provides an overview of the current state of cybersecurity threats and trends in Europe, as well as globally. It analyzes the most significant cybersecurity incidents that occurred during the previous year and identifies emerging threats and vulnerabilities that could impact individuals, organizations, and critical infrastructure.
	
	The report also includes recommendations for improving cybersecurity such as best practices for risk management and incident response. It is intended for a broad audience, including policymakers, IT professionals, and the general public, to increase awareness of the current cybersecurity landscape and help stakeholders make informed decisions about how to protect their digital assets.
	
	The ENISA Threat Report is an important resource for anyone interested in understanding the current state of cybersecurity and how to mitigate potential risks and is beneficial to establish which important skills are required to mitigate the exposure of cyberattacks.
	
	\subsection{NIST Cybersecurity Framework}
	\label{sec:NIST}
	The NIST Cybersecurity Framework~\cite{stine2014framework} is a set of guidelines, standards and best practices developed by the National Institute of Standards and Technology (NIST) to help organizations manage and reduce cybersecurity risks. The framework consists of five core functions: Identify, Protect, Detect, Respond, and Recover. Each function contains a set of categories and subcategories that provide detailed guidance on specific actions that organizations can take to enhance their cybersecurity posture. The framework is widely recognized and used by government agencies, private companies, and organizations of all sizes to improve their cybersecurity practices and mitigate the risk of cyberthreats.
	
	The framework was first released by NIST in 2014, and it has since been widely adopted by organizations in various industries, including healthcare, finance, and energy. It is designed to be flexible and adaptable to the unique needs and risk profiles of each organization, regardless of size or sector.
	
	The five core functions of the framework are as follows:
	\begin{itemize}
		\item Identify: This function involves developing an understanding of the organization's systems, assets, data, and risks. It includes activities such as inventorying hardware and software assets, identifying vulnerabilities and threats, and assessing the potential impact of cybersecurity incidents.
		
		\item Protect: This function focuses on implementing safeguards to protect against cyber threats. It includes activities such as access control, data encryption, and security awareness training for employees.
		
		\item Detect: This function involves identifying cybersecurity incidents as quickly as possible. It includes activities such as continuous monitoring, anomaly detection, and incident response planning.
		
		\item Respond: This function involves taking immediate action to contain and mitigate the effects of cybersecurity incidents. It includes activities such as incident response, business continuity planning, and disaster recovery.
		
		\item Recover: This function involves restoring normal operations after a cybersecurity incident. It includes activities such as system recovery, damage assessment, and post-incident review.
	\end{itemize}
	
	The NIST Cybersecurity Framework further defines a list of skills which are essential for a better protection from cyber risks, for both developers and general users of computer systems as the following:  
	\begin{itemize}
		\item Preventing malware via non-trustworthy websites
		\item Preventing malware via email phishing
		\item Preventing Personal Identifiable Information theft via access to non-trustworthy websites
		\item Preventing Personal Identifiable Information theft via email phishing
		\item Preventing Personal Identifiable Information via social media
		\item Preventing information system compromise via USB or storage device exploitation
		\item Preventing unauthorized information system access via password exploitation
	\end{itemize}
	
	The framework is intended to be used as a tool for improving an organization's cybersecurity practices, rather than as a one-size-fits-all solution. It is a living document that can be updated and customized over time to reflect changes in technology, threats, and business needs.
	
	\subsection{Phishing in Organizations: Findings from a Large-Scale and Long-Term Study}
	The research project \emph{Phishing in Organizations: Findings from a Large-Scale and Long-Term Study}~\cite{lain2022phishing} ran a large scale experiment in a company where more than 14'000 employees participated. The goals were to understand if certain employees are likelier to fall for phishing than others, how the overall vulnerability to phishing of the company evolves over time, how effective phishing warning and training is and how effectively crowd-sourced phishing detection is applied in companies. The study confirms previous findings that both age and computer skills correlate with being susceptible to phishing. Further it was revealed that the most vulnerable employees were those that use computers daily for repetitive tasks using specialized software only. A more concerning finding of the study was that simulated phishing exercises and voluntary training did not just fail to improve resilience but, rather contradictory, made them more prone to fall for phishing attacks. The severity of this finding is even worse when combined with the observation that a longer exposure to phishing attacks may lead to a higher portion of employees being susceptible for it. On the positive side it was demonstrated that crowd-sourced phishing detection can be efficient in large organizations. 
	
	\subsection{Cyber Security Training A Survey of Serious Games in Cyber Security}
	In the paper \emph{Cyber security training a survey of serious games in cyber security}~\cite{Tioh2017CysecTraining} a survey of academic and non academic serious games that focus on cybersecurity was conducted. The study emphasizes that while a lot of effort has been put into technical security, drastically less attention has been put into how to better educate users, which are considered to be the weakest link in the security chain. It is further stated that effective large scale measures rather target a technical audience while little is done to overcome the perception that cybersecurity is for tech savvy people only. Compared to hands on training methods game based learning allows students, or even encourages, to make mistakes in a risk-free environment and learn from them~\cite{Squire2003VideoGamesInEducation}. Further it is argued that well designed serious games can retain all advantages of traditional and hands-on training while remaining low-cost. Based on the survey it is concluded that serious games for cybersecurity seem promising but there is a lack of proper evaluation of these games and therefore no conclusive answer on the effectiveness can be given.
	
	
	\section{Research Design}
	\label{sec:Research_Design}
	The goal of our cyberattack simulator is to improve the knowledge about cyberattacks in the general society. Therefore, we designed a study to question players of our game before and after playing to understand the impact of the game on the knowledge about cyberattacks. The related questionnaire is organized in a structured form, so asking either \emph{yes or no} or range questions with a range from one to five. The questions have a main focus on the phishing part of the game, since we assume that most people have been in some contact with phishing~\cite{alabdan2020phishing}. We distributed our questionnaire to bachelor students of two different research institutes as well as to further people randomly. 
	The first survey before the game focused on the existing knowledge of the participants in the field of cyberattacks. The second survey after the game verified whether the knowledge about cyberattacks improved by playing the game, thus it consists of almost completely the same questions. 
	The surveys were done anonymously, and we did not track the answers of individual participants from before and after the game. 
	Our study only analyzed the short term effects, as we did not perform any assessment a longer period after playing the game (e.g., half a year or a year later). 

	\section{Game Design}
	\label{sect:GameDesign}
	In this section we describe the game design of our cyberattack simulator. We focus on the most important part of the game -- the scenarios. 
	Later, we will discuss the actual game design such as the graphics and style.
	
	The central goal of the game is to give the players, regardless of their technical knowledge, a better understanding on how cyber criminals perform cyberattacks and how they approach targets by letting them perform the attacks themselves. Attack centric games are not uncommon to teach cyber security skills and are often encountered in the form of Capture the Flag (CTF) games. The drawback is that they usually either require extensive technical knowledge or time to play. Our serious game approach aims to make cybersecurity games more accessible to a wider audience and allow them to explore freely, without having to fear making mistakes. The game is divided into different scenarios, each corresponding to real world attacks that bad actors use. By interacting with the character \emph{Nate} the player can choose which scenario to play first. After choosing a scenario the player moves their character to a computer in the game from which the chosen attack is launched. The following subsections will first provide a mapping from the NIST framework to our scenarios and then introduce and explain the currently available scenarios for players to explore.
	
	\subsection{Mapping Cybersecurity Skills to Game Scenarios}
	
	We used the NIST framework (see section~\ref{sec:NIST}) as basis for our scenarios and summarizing the different skills required to successfully face cyberattacks into four categories. 
	
	We start by summarizing the skills into categories.
	\begin{enumerate}
		\item The first scenarios is add \emph{Phishing/Social Engineering} where the skills \emph{Preventing malware via email phishing}, \emph{Preventing Personal Identifiable Information theft via email phishing}, \emph{Preventing Personal Identifiable Information theft via access to non-trustworthy websites} and \emph{Preventing Personal Identifiable Information via social media} are covered. 
		The scenario will walk a player through a typical phishing attack to understand the way an attacker thinks and to educate them on the concepts they should recognize in phishing mails. 
		\item The second category is more of a category for developers. It covers the \emph{Preventing unauthorized information system access via password exploitation} skill. The scenario walks the player through a situation where an attacker can steal information from a non secure website. 
		\item The third category is about Metasploit~\cite{kennedy2011metasploit}, which is an attack framework for penetration testing but could also can be abused by attackers, and covers the importance of patching and updating critical systems. This is also a more developer and system engineer focused scenario and relies on recommendations from the ENISA thread landscape (see section \ref{sec:ENISA}) and covers the skill \emph{Preventing malware via non-trustworthy websites}. 
		\item The fourth and last category is about \emph{bad USB keys}. It covers the skill \emph{Preventing information system compromise via USB or storage device exploitation} and walks the player through how attackers could obtain malware from the darknet and use USB keys to find a way into a company network.
	\end{enumerate}
	
	For each category a scenario is created. In the following section all scenarios are described in more detail. 
	Currently, the scenarios are sample scenarios we created based on the information from the NIST framework and the ENISA threat landscape to cover and evaluate the base skills. In the future more scenarios will be developed to cover more advance attack scenarios.

	\subsection{Social Engineering and Phishing}
	The term social engineering describes the act of deceiving an individual into revealing sensitive information, obtaining unauthorized access, or committing fraud by associating with the individual to gain confidence and trust.\footnote{\url{https://nvlpubs.nist.gov/nistpubs/SpecialPublications/NIST.SP.800-63-3.pdf}} In order to convince victims to provide information or services social engineers rely heavily on the six principles of influence~\cite{Wissler2002TheSO}. Based on these principles social engineers use the following with slight adaptions: authority, intimidation, consensus/social proof, scarcity, urgency and familiarity/liking. Commonly the attacker engages the target over email which is known as phishing in the general case and spear-phishing if it is targeted to a specific organization or individual. The Verizon Data Breach Investigations Report \footnote{\url{https://www.verizon.com/business/resources/reports/dbir/}} states that more than 60\% of the breaches in Europe, the Middle East and Africa included a social engineering component and despite awareness campaigns and exercises many still fall for phishing mails. Social engineering is the most common entry point for attackers and the easiest to avoid if employees are properly educated. Further it requires very little technical knowledge to defend against phishing and therefore anyone can and must be properly trained against it. 
	
	In our game the player begins the social engineering attack by using social media in order to gather information about the target. After finding a business email address on social media the player proceeds to check if the email has been part of a known data breach where passwords have been leaked. As the email has not been part of a leak the player receives the option to start a phishing attack. The game proceeds to guide the player step by step through a simplified but real phishing attack using SocialPhish\footnote{\url{https://github.com/xHak9x/SocialPhish}}. More complicated commands and searches are replaced through clicks and specific text inputs to be more user friendly. In order to get the login credentials of a target the game makes use of the principle of urgency by sending a fake email from Facebook that tells the target that his account is about to expire. The target falls for the phishing email and the phished credentials are displayed on the screen.
	
	\subsection{SQL Injection}
	In 2021 Injection based attacks were placed third in OWASP's Top 10 ranking~\cite{owasptop10:2021}, a standard awareness document for developers describing the Top 10 web application security risks. In the previous ranking of 2017  Injections were first place. SQL Injections are a particular type of an injection based attack where databases are targeted. By injecting SQL statements into input data an attacker tries to manipulate the queries such that sensitive data is manipulated or revealed. Further consequences of a successful attack are executing administrator commands in the Database Management System, try to steal content of files present in the database and in some cases even to execute commands on the underlying operating system. Software is susceptible to SQL Injections if it interacts with an SQL database, takes data from an untrusted source, such as user input on a website, and dynamically uses that data to construct queries~\cite{halfond2006classification}. 
	
	As this is a more technical and advanced concept our game only briefly introduces the concept of SQL Injections by walking the player through an attack and explaining the underlying concepts. The Injection allows the player to bypass a login and access the test server in order to search through the files. By looking into the various folders (and based on their contents) possible follow up actions, as well as their intentions, are explained.
	
	\subsection{Metasploit}
	The Metasploit Framework~\cite{stine2014framework} is a tool for creating and executing exploits against targeted machines. At its core it is a collection of known vulnerabilities and payloads that is meant to help penetration testers to better test systems and always having an up to date catalog of attack options. When performing an attack with the framework it supplies a skeletons of known exploits and lets the user manually set specific targets and options. Further the framework provides Meterpreter, a payload that allows to take over control of the victims system. Both the Metasploit Framework and Meterpreter are publicly available and therefore not only used by security experts to attack systems in a legal manner but also by attackers performing illegal actions. 
	
	The scenario begins with the player executing a network scan using the tool Nmap~\cite{nmap}, a popular security scanner that scans networks for hosts and which services are running on them. After performing the initial scan the player is then guided through the typical setup of an exploit in the Metasploit Framework and its configuration to attack a specific target with a selected payload. After typing \emph{run} the exploit executes successfully and a Meterpreter session is started on the targeted device. This session is then used to find and download sensitive data. The scenario ends with a note that at this point the system is severely exposed and that an attacker can now perform various malicious actions, e.g., the installing of malware on the targeted host.
	
	\subsection{Bad USB}
	Bad USB also known under the name of Rubber Ducky, a bad USB manufactured by Hak5\footnote{\url{https://shop.hak5.org/}}, is an attack where modified USB sticks are used to attack systems. These USB sticks usually contain additional programmable microcontroller that allow the attacker to program it~\cite{vouteva2015feasibility}. Once plugged into a system the bad USB emulates a keyboard and is directly recognized from the targeted system as such. The device then initiates keystrokes and issues commands to the targeted system. Such commands can be opening a terminal and downloading ransomware, find and steal sensitive data or harmlessly prank users by turning the targeted system off. In order to be successful the attacker only needs to plug in the USB into the target system or convince someone else to do so. There are also large fashion attacks where attackers send USB devices as gifts to employees of a targeted company. By sending bad USB's to hundreds of employees there is a high chance that one of them will use it in the near future on a company device, giving the attacker an entry point to further attack the target~\cite{scanavez2021bad}.
	
	To be able to start this scenario the player first needs to find the USB stick placed on a desk in the room. The game then explains how a bad USB attack works and gives the player a choice on how to proceed. The first option is to flash it with a Zero Day vulnerability, a vulnerability that is yet not known to the producers of software. The game illustrates how cyber criminals can get hold of such vulnerabilities by guiding the player to the darknet to purchase such an exploit. It also explains how cyber criminals remain anonymous while using online black markets. After purchasing the Zero Day the scenario ends in letting the player choose how to label the USB stick with a wording that would motivate employees finding it to plug it in into their computer.
	
	The second option is less malicious and includes a batch script that opens Microsoft Word  5 times. It is further explained that this script can be linked to an image in order to increase the likelihood that someone accidentally executes it. While in this scenario a harmless example is used, the script is interchangeable and can contain far more malicious contents. This option ends in the player proceeding to take the aforementioned Zero Day option. 
	
	\subsection{Game Implementation}
\begin{figure}
		\centering
		\includegraphics[width=0.5\linewidth]{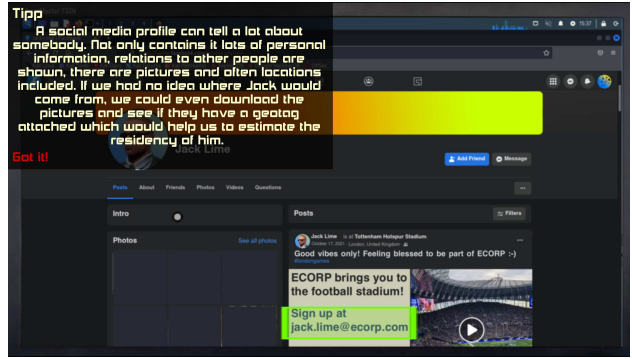}
		\caption{Screenshot of the implemented phishing scenario.}
		\label{fig:phishing_game_scenario}
	\end{figure}

	The design of the game has been influenced by Cybersiege~\cite{Thompson2011ActiveLW, Thompson2014CyberCIEGESD, Thompson2016CyberCIEGEA} which led to the design of a small narrative that is followed by different scenarios showcasing real cyberattacks. The game is 3D which allows the player to walk freely around the room and explore it. It has been developed by using Adventure Creator\footnote{\url{https://adventurecreator.org/}}, a plugin for the Unity\footnote{\url{https://unity.com}} game engine that allows to develop games with visual scripting. When the player interacts with the in-game-computer the scene switches to a pseudo 2D environment only showing the screen. In order to be able to run in a browser with WebGL the game was heavily compressed to a size of 190MB. Figure~\ref{fig:phishing_game_scenario} shows a screenshot of the phishing scenario in the game where the game informs the player about the potential risks of posting sensitive information on social media.
	
	\section{Evaluation}
	\label{sect:eval}
\begin{figure}[h]
		\centering
		\includegraphics[width=0.45\linewidth]{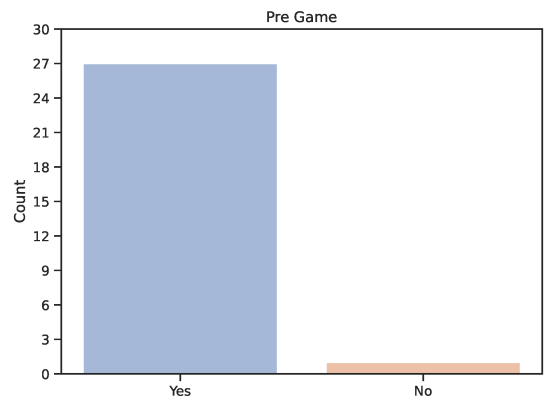}
		\caption{Answers to the question: Do you know what a phishing mail is?}
		\label{fig:phishing_pregame}
	\end{figure}

	In order to evaluate our game we conducted a short term survey (see section~\ref{sec:Research_Design}) where the participants were asked to fill out a form before and after the game. As usual with pre and post surveys we received more pre than post results. We did not opt to add personal identifiers to each form in order to preserve the privacy of the participants. This made it impossible to link pregame forms to postgame forms. In figure~\ref{fig:phishing_pregame} we see that almost all participants knew what a phishing mail is. 
	
\begin{figure}
		\centering
		\includegraphics[width=0.7\linewidth]{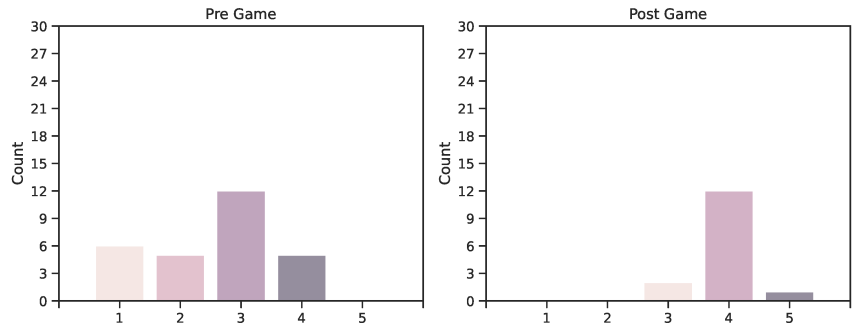}
		\caption{Answers to the question: Do you know how a phishing attack is rolled out?}
		\label{fig:phishing_attack_rolled_out}
	\end{figure}

\begin{figure}
		\centering
		\includegraphics[width=0.7\linewidth]{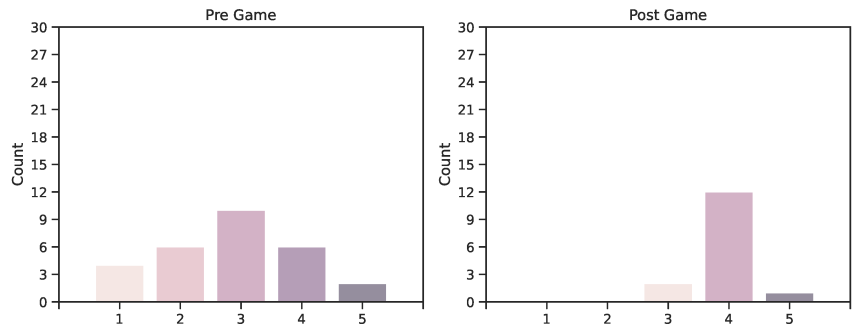}
		\caption{Answers to the question: Do you know, how a phishing attacker acquires your email address?}
		\label{fig:phishing_aquire_email}
	\end{figure}

\begin{figure}
		\centering
		\includegraphics[width=0.4\linewidth]{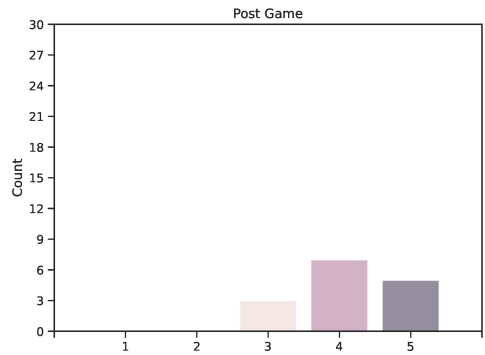}
		\caption{Answers to the question: Do you think the game "The get in" helped you to develop a better understanding about cyber attacks and cyberrisks?}
		\label{fig:better_understanding}
	\end{figure}

	But as seen in figure~\ref{fig:phishing_attack_rolled_out} most of the participants had little to no knowledge on how a phishing attack is rolled out before playing our game. In the post game results we can see a clear improvement of the understanding on how a phishing attack is rolled out. Despite the imbalance of pre and post results we conclude from the absolute numbers that the participants gained a better understanding from playing our game. As shown in figure~\ref{fig:phishing_aquire_email}, we also asked the participants if they know how phishing attackers can acquire their email. Also here the awareness on how attackers can acquire emails of victims increased. At the end of the survey we asked the participants whether they in general learned more about cyber attacks and risks through our game. From figure~\ref{fig:better_understanding} it can be concluded, that on average, the game has led to an improved understanding of cyber attacks and cyberrisks for the players of the game.

	\section{Conclusion \& Future Work}
	\label{sect:conclusion}
	In this paper we highlighted common cyber threats for employees in their day to day business and the related potential consequences. Further we emphasized that for phishing and social engineering, the most common entry points for cyber criminals, current prevention measures taken actually make employees less robust against these attacks~\cite{lain2022phishing}. In order to increase awareness and understanding of common cyber threats we proposed a serious game instead of awareness campaigns. In this game the player takes the role of the attacker and gets insights on how attackers approach their targets and exploit common pitfalls. Attack based games are very popular in the cybersecurity world and are mostly encountered in the form of CTFs. These gameified scenarios, that usually correspond to real scenarios, have a huge learning effect but often require a deep technical understanding which makes them less accessible for a wider audience. Our approach is an interactive process in which the player is guided through different attack scenarios to make it more accessible. In each step of the attack the reasoning as well as options for prevention are is explained. 
	
	Based on our evaluation in section~\ref{sect:eval} we conclude that our game has a positive short term effect on increasing cybersecurity awareness. Furthermore, almost all participants that played the game stated that they enjoyed doing so. In order to mitigate the issue of the inequality of pre- and post game responses we could add a randomized session identifier to the pregame form that the player then should use for the game and the post form. With these measures we could filter out partial results and draw better conclusions. It is important to notice that we did not conduct a phishing awareness campaign before and after the participants played our game to verify the educational effects of the game in practice. The questionnaire itself also focuses more on short term results. As pinpointed in~\cite{Tioh2017CysecTraining} there is currently no long term evaluation on the effect of serious games in a cyber security setting. We also leave that open as future work. 
	
	The game itself is currently in a development state and needs further refinement. Further it is questionable if the 3D setting adds any particular benefits other than mimicking AAA games. Especially for people that are not used to playing 3D games the more complex movement controls can potentially distract from the actual objective to highlight cyber risks. As the main focus of the game lies on scenarios which play on a computer (i.e. a 2D setting), the 3D graphics add more complexity than benefit for gameplay. The time spent on the added complexity of developing 3D games would be better invested in creating more sophisticated 2D scenarios, adding different difficulty options for each scenario or better image quality. 
	
	\label{sect:bib}
	\bibliographystyle{unsrt} 
	\bibliography{easychair}
	
\end{document}